\documentclass{article}
\usepackage{spconf,amsmath,graphicx}
\usepackage{color}
\usepackage{hyperref}

\title{BOOTSTRAPPING NON-PARALLEL VOICE CONVERSION FROM SPEAKER-ADAPTIVE TEXT-TO-SPEECH}

\name{Hieu-Thi Luong\textsuperscript{1,2}, Junichi Yamagishi\textsuperscript{1,2,3}\thanks{This work was partially supported by a JST CREST Grant (JPMJCR18A6, VoicePersonae project), Japan, and MEXT KAKENHI Grants (16H06302, 17H04687, 18H04120, 18H04112, 18KT0051), Japan.}}
\address{\textsuperscript{1}SOKENDAI (The Graduate University for Advanced Studies), Kanagawa, Japan \\
\textsuperscript{2}National Institute of Informatics, Tokyo, Japan\quad{}\quad{}\textsuperscript{3}The University of Edinburgh, Edinburgh, UK}

\begin{document}
\maketitle
\begin{abstract}
Voice conversion (VC) and text-to-speech (TTS) are two tasks that share a similar objective, generating speech with a target voice. However, they are usually developed independently under vastly different frameworks. In this paper, we propose a methodology to bootstrap a VC system from a pretrained speaker-adaptive TTS model and unify the techniques as well as the interpretations of these two tasks. Moreover by offloading the heavy data demand to the training stage of the TTS model, our VC system can be built using a small amount of target speaker speech data. It also opens up the possibility of using speech in a foreign unseen language to build the system. Our subjective evaluations show that the proposed framework is able to not only achieve competitive performance in the standard intra-language scenario but also adapt and convert using speech utterances in an unseen language.
\end{abstract}
\section{Introduction}
\label{sec:intro}

The voice conversion (VC) system is used to convert speech of a certain speaker to speech of a desired target while retaining the linguistic content \cite{toda2007voice}.
VC and text-to-speech (TTS) systems are two different systems functioning under different scenarios, but they share a common goal: generating speech with a target voice. If we define the TTS system as a synthesis system generating speech using linguistic instructions (in an abstract sense) obtained from written text \cite{taylor2009text}, then the VC system can be defined as a synthesis system generating speech using linguistic instructions extracted from a reference utterance.
This similarity in objective but difference in operation context makes VC and TTS complement each other and have their unique role in a spoken dialogue system.
More specifically, as text input is easy to create and modify, TTS is capable of generating a large amount of speech automatically and cheaply. However it is difficult to generate speech when the desired linguistic instructions cannot be represented in the expected written form (e.g. when text is written in foreign languages). On the other hand, speech used as a reference input for VC is more time-consuming and expensive to produce, but the system can be straightforwardly extended to an unseen language.

A VC system is usually classified as parallel or non-parallel depending on the nature of the data available for development. 
The parallel VC system is developed using a parallel corpus containing pairs of utterance spoken by a deterministic source and target speakers. The parallel speech utterances of source and target are aligned using dynamic time warping (DTW) to create the training set. By using this training set, a mapping function is formulated to transform the acoustic features of one speaker to another \cite{abe1990voice}.
Many methods have been proposed to model this transformation for parallel VC systems \cite{toda2007voice,takashima2013exemplar,chen2014voice}.
Parallel VC is developed with just data of source and target speakers which is one of its advantages.
Acquisition of parallel speech data requires a lot more planning and preparation than acquisition of non-parallel speech, which in turn makes the former more expensive to develop, especially when we want a VC system with the voice of a particular speaker.
Therefore, many approaches have been proposed to develop VC systems using a non-parallel corpus \cite{erro2009inca,chen2003voice}. For the conventional Gaussian mixture model (GMM) approach, non-parallel VC can be adapted from a pretrained parallel VC in the model space using the maximum a posterior (MAP) method \cite{chen2003voice,lee2006map} or as interpolation between multiple parallel models \cite{toda2006eigenvoice}. For recent neural network approaches, a non-parallel VC can be trained by directly using an intermediate linguistic representation extracted from an automatic speech recognition (ASR) model \cite{sun2016phonetic,miyoshi2017voice} or by indirectly encouraging the network to disentangle linguistic information from the speaker characteristics using methods like variational autoencoder (VAE) \cite{hsu2016voice}, generative adversarial networks (GAN) \cite{kaneko2017parallel,fang2018high} or some other techniques \cite{tjandra2019vqvae}.
For both parallel and non-parallel VC, the systems usually change the voice but are unable to change the duration of the utterance. Many recent researches focused on converting speaking rate along with voices by using sequence-to-sequence models \cite{miyoshi2017voice,yeh2018rhythm,tanaka2019atts2s,zhang2019non}, as speaking rate is also a speaker characteristic.

Recently Luong et al. proposed a speaker-adaptive TTS model that could perform speaker adaptation with untranscribed speech using backpropagation algorithm \cite{luong2019unified}. We find that the proposed system can potentially be used for developing non-parallel VC as well. 
In this paper, we introduce a framework for creating a VC system using the untranscribed speech of a target speaker by bootstrapping from the speaker-adaptive TTS model proposed by Luong et al. \cite{luong2019unified}. Furthermore, we investigate the performance of our framework when it adapts and convert using speech utterances of an unseen language.
The rest of paper is organized as follows. Section \ref{sec:architecture} describes our framework, Section \ref{sec:unseen} introduces different scenarios in which a VC can operate in an unseen language and their practical applications, Section \ref{sec:experiment} describes the experiment setup, Section \ref{sec:evaluation} presents subjective results, and Section \ref{sec:conclusion} concludes our findings.

\section{Bootstrapping voice conversion from speaker-adaptive TTS}
\label{sec:architecture}
\subsection{Multimodal architecture for speaker-adaptive TTS}
\label{sec:ttsmodel}

\begin{figure}[t]
  \centering
  \includegraphics[height=0.9\linewidth]{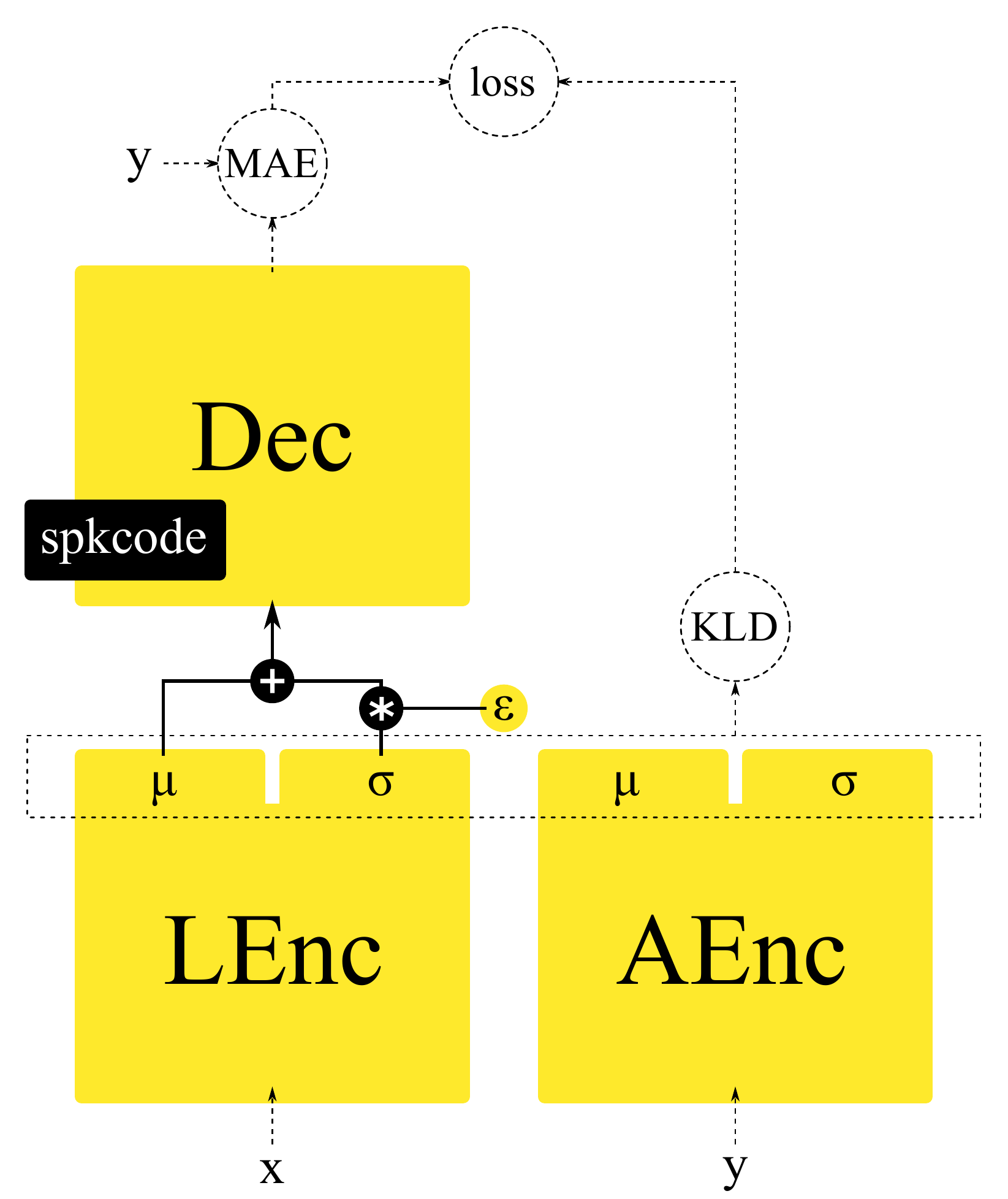}
  \vspace{-3mm}
  \caption{Training the initial speaker-adaptive TTS model.}
  \label{fig:proposaltts}
  \vspace{-3mm}
\end{figure}

\begin{figure}[t]
  \centering
  \includegraphics[height=0.9\linewidth]{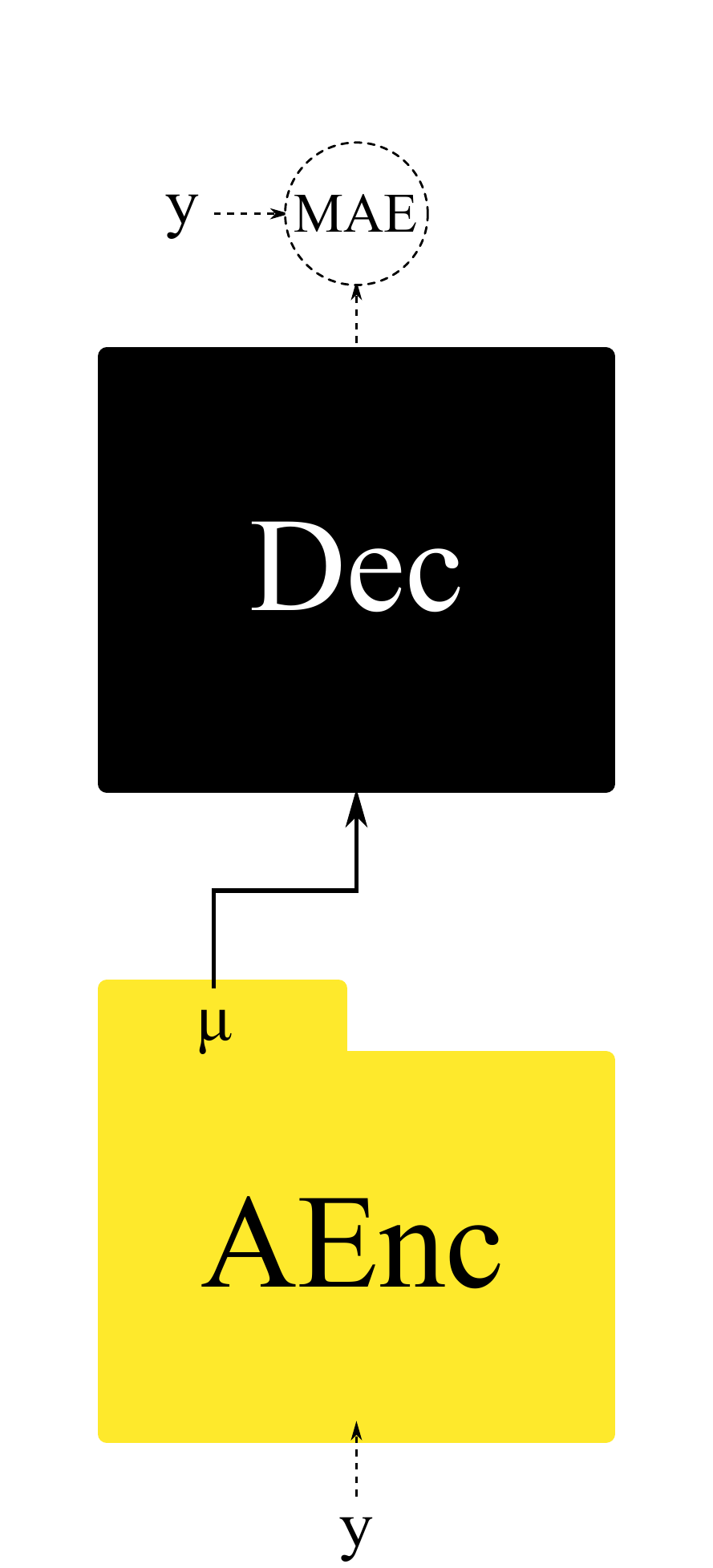}
  \vspace{-3mm}
  \caption{Transferring to VC and adapting to the target speaker.}
  \label{fig:proposalvc}
  \vspace{-3mm}
\end{figure}

We first summarize the speaker-adaptive TTS proposed in \cite{luong2019unified}. A conventional TTS model is essentially a function that transforms linguistic features $\boldsymbol{x}$, extracted from a given text, to the acoustic features $\boldsymbol{y}$, which could be used to synthesize speech waveform. A neural TTS model is a TTS model that uses neural networks as the approximation function for the $\boldsymbol{x}$ to $\boldsymbol{y}$ transformation. Neural networks are trained with backpropagation algorithm and can also be adapted to a new unseen domain with backpropagation when the adaptation data is labeled. In the context of TTS, an acoustic model can easily be adapted to voices of unseen speakers if the adaptation data is transcribed speech but it is not as straightforward when the adaptation is untranscribed. Many techniques have been proposed to perform adaptation with untranscribed speech by avoiding backpropagation and extracting a certain speaker representation from an external system instead \cite{jia2018transfer,nachmani2018fitting}. However, this forward-based approach limited the performance of the adaptation process \cite{luong2019unified}.

Luong et al. \cite{luong2019unified} proposed backpropagation-based adaptation method using untranscribed speech by introducing a latent variable $\boldsymbol{z}$, namely latent linguistic embedding (LLE), which presumably contains no information about the speakers. The conventional acoustic model then splits into a speaker-dependent (SD) acoustic decoder $Dec$ and a speaker-independent (SI) linguistic encoder $LEnc$. The acoustic decoder is defined by its SI parameter $\theta^{\textrm{core}}$ and SD parameter $\theta^{\textrm{spk},(k)}$, one for each $k$-th speaker in the training set while the linguistic encoder is defined by its SI parameter $\phi^L$. The TTS model can be described as follows:
\begin{align}
    \boldsymbol{z}^L &\sim LEnc(\boldsymbol{x};\phi^L) = p(\boldsymbol{z}|\boldsymbol{x})\\
    \boldsymbol{\tilde{y}}^L &= Dec(\boldsymbol{z}^L;\theta^{\text{core}},\theta^{\text{spk},(k)})
\end{align}
With this new factorized model, if we want to adapt the acoustic decoder to unseen speakers using speech, we train an auxiliary acoustic encoder $AEnc$ (defined by its SI parameter $\phi^A$) to use as a substitute for the linguistic encoder:
\begin{align}
\label{equa:zA}
    \boldsymbol{z}^A &\sim AEnc(\boldsymbol{y};\phi^A) = q(\boldsymbol{z}|\boldsymbol{y})\\
    \boldsymbol{\tilde{y}}^A &= Dec(\boldsymbol{z}^A;\theta^{\text{core}},\theta^{\text{spk},(k)})
\end{align}
The acoustic encoder is only used to perform the unsupervised speaker adaptation in \cite{luong2019unified}; however the network created by stacking the acoustic encoder and acoustic decoder is essentially a speech-to-speech stack that potentially could be used as a VC system. This is the motivation for us to develop a VC framework based on this model.

\subsection{Latent Linguistic Embedding for many-to-one voice conversion}
\label{sec:vcmodel}

By using the speaker-adaptive TTS model described above, we propose a three-step framework to develop a VC system for a target speaker:

\textbf{1) train the initial TTS model}: The first step is to train the TTS model, it is identical to the \textit{training mode} defined in Section IV-B in \cite{luong2019unified}. The data required for this step is a multi-speaker transcribed speech corpus. In this step, we need to jointly train all modules using a deterministic loss function:
\begin{equation}
\label{equa:train}
    \textrm{loss}_{train} = \textrm{loss}_{main} + \beta \; \textrm{loss}_{tie}
\end{equation}
$\textrm{loss}_{main}$ is the distortion between output of the TTS stack and the natural acoustic features, and $\textrm{loss}_{tie}$ is the KL divergence (KLD) between the output of the linguistic encoder and output of the acoustic encoder.
The setup is almost the same as described in \cite{luong2019unified}. One small difference is that we use the mean absolute error (MAE) as the distortion function instead of the mean square error as shown in Fig. \ref{fig:proposaltts}. In this step, the acoustic encoder is trained to transform acoustic features to the linguistic representation LLE while the acoustic decoder is trained to become a good initial model for speaker adaptation.

\textbf{2) transfer to VC and adapt to the target}: For building a VC system we only need the acoustic encoder and the acoustic decoder trained previously; the linguistic encoder is discarded as it is no longer needed. The acoustic encoder and decoder make a complete speech-to-speech network. 
To have a VC system with a particular voice we adapt the acoustic decoder using the available speech data of the target. This is identical to the \textit{unsupervised adaptation mode} described in Section IV-B in \cite{luong2019unified}. More specifically, we removed all SD parameters $\theta^{\textrm{spk},(k)}$ and then fine-tuned remaining parameter $\theta^{\textrm{core},(r)}$ to the $r$-th target speaker by minimizing the loss function:
\begin{equation}
    \textrm{loss}_{adapt} = L_{MAE}(\boldsymbol{\tilde{y}}^A,\boldsymbol{y})
\end{equation}
Slightly different to \cite{luong2019unified}, we removed the standard deviation $\sigma$ from the acoustic encoder as illustrated in Fig. \ref{fig:proposalvc}. Simply speaking instead of sampling $\boldsymbol{z}^A$ from a distribution (Equation \ref{equa:zA}), $\boldsymbol{z}^A$ will always take the mean value instead. We found this change slightly improves the performance of the adapted model in the converting step.

\textbf{3) converting speech of arbitrary source speakers}: The adapted VC system then could be used to convert speech of arbitrary source speakers to the target. Our system does not need to train on or adapt to the speech data of source speakers, although doing so might help the performance.

\subsection{Related works}
\label{subsec:relatedworks}

While the original VC is narrowly defined as a function that directly maps acoustic units of a source to a target speaker \cite{abe1990voice,stylianou1998continuous}, recent VC systems have usually been modeled with an intermediate linguistic representation, either explicitly \cite{sun2016phonetic} or implicitly \cite{kameoka2018acvae} due to the rapid development of representation learning and disentanglement ability of neural networks.
This approach has further reduced the theoretical difference between VC and TTS. Motivated by the same observation, Mingyang Zhang et al. \cite{zhang2019joint} proposed a method to jointly train a TTS and a parallel VC system using a shared sequence-to-sequence model. The proposed framework improves the stability of the generated speech but does not add any benefits in term of data efficiency, as parallel speech data of the source and target speakers is still a requirement. Recently, Jing-Xuan Zhang et al. \cite{zhang2019non} have introduced a non-parallel sequence-to-sequence voice conversion system that has a procedure that is similar to our: the initial model is trained with a TTS-like network to help disentangle linguistic representation and the model then adapted to the source and target speakers. However, the adaptation step of the framework proposed in \cite{zhang2019non} requires both source and target speaker speech as well as their transcripts, which increases the data demand for building a VC system with a particular voice.

Despite being different in motivation and procedure, our framework has the same data efficiency as the models using phonetic posteriorgrams (PPG) proposed by Sun et al. \cite{sun2016phonetic}. The LLEs in our setup play a similar role to the PPGs, but are very different in terms of characteristics. While PPGs are the direct output of an ASR network with an explicit interpretation, LLEs are latent variables and guided by a TTS network. The efficiency in terms of data usage is not only lowering the requirement for building a VC system but also opening up the possibility of solving other interesting tasks. Creating and using the VC system with a low-resource unseen language is one example \cite{sun2016personalized}.

\section{Using voice conversion system in an unseen language}
\label{sec:unseen}

The proposed three-step framework have an interesting characteristic which is the asymmetric nature in terms of data requirements for each step. While the TTS training step requires a transcribed multi-speaker corpus, the adaptation step only requires a small amount of untranscribed speech from the target speaker. Moreover the model does not need to train on data of source speakers and hypothetically can convert utterances of arbitrary speakers out of the box. We could take advantage of these characteristics to build VC systems for low-resource languages.
For our experiments, we define the abundant language as language with transcribed speech corpus and used to train the initial TTS model (seen); in our case it is English (E). The low-resource language is a language whose data does not include transcripts and is not used in the TTS step (unseen); in our case it is Japanese (J).
We first define multiple scenarios in which our framework can interact with an unseen language. These scenarios are dictated by the language used in the adaptation and conversion steps:

\begin{figure}
  \centering
  \includegraphics[width=0.9\linewidth]{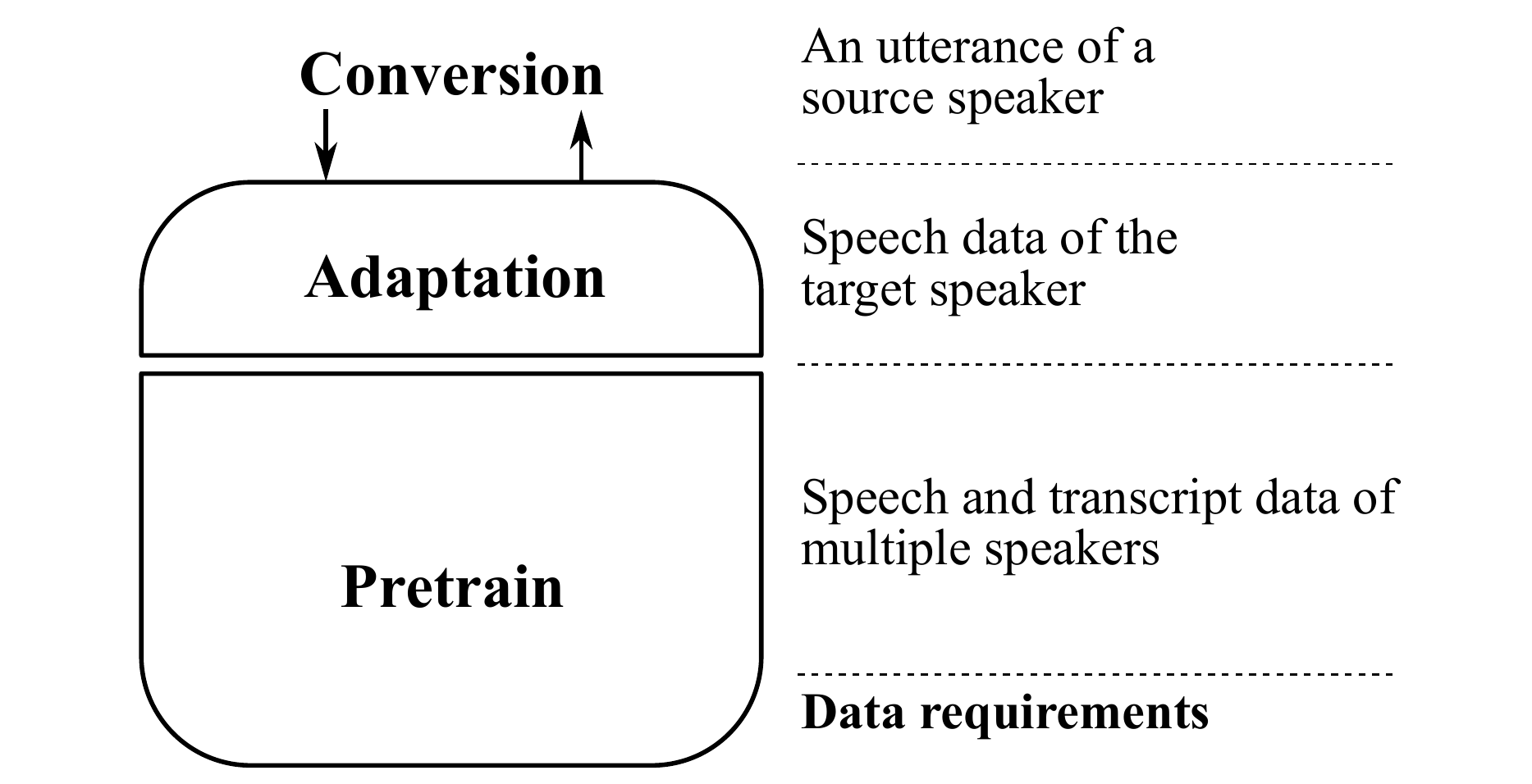}
  \caption{The asymmetric nature of data requirements for each step of the proposed framework.}
\label{fig:langinfo}
\vspace{-3mm}
\end{figure}

\textbf{(a) EE-E}:  We adapt the pretrained English model to an English speaker and using it to convert English utterances. Generally speaking the entire framework from start to finish operates within a single language. This is the typical intra-language scenario of a voice conversion system.

\textbf{(b) EE-J}: We adapt the pretrained English model to an English speaker but use it to convert Japanese utterances. Generally speaking, the VC system is used to generate speech of linguistic instructions that have not been seen. The ability to generate speech for content that is difficult to represent in the expected written form (a foreign language in this case) is one advantage of VC over TTS. This scenario is referred as cross-language voice conversion \cite{abe1991statistical,mashimo2001evaluation}.

\textbf{(c) EJ-E}: We adapt the pretrained English model to a Japanese speaker and use it to convert English utterances. In this scenario we want to build a VC system in the abundant language; however, the speech data of the target speaker is only available in a low-resource unseen language. This is sometime referred to as cross-lingual voice conversion \cite{zhou2019cross,rallabandi2019approach}, however we call it cross-language speaker adaption to distinguish it from EE-J. Unlike other scenarios involving the unseen language, cross-language speaker adaptation is relevant for both VC \cite{zhou2019cross} and TTS \cite{wu2009state,sun2016personalized}. Even though it is not evaluated in this paper, this scenario is also the unsupervised cross-language speaker adaptation scenario of the TTS system proposed in \cite{luong2019unified}.

\textbf{(d) EJ-J}: We adapt the pretrained English model to a Japanese speaker and use it to convert Japanese utterances. In this scenario we essentially bootstrap a VC system for a low-resource language from a pretrained model of an abundant one. The written form of the target language is not used in the training, the adaptation or the conversion stages. This scenario is sometime referred as text-to-speech without text, which is the main topic of the Zero Resource Speech Challenge 2019 \cite{dunbar2019zero}. Even though our scenario has the same objective as the challenge, the approach is a little different. The participant of the challenge are encouraged to develop intra-language unsupervised unit discovery methods, which are more difficult \cite{feng2019combining,liu2019unsupervised,eloff2019unsupervised}. Our framework is bootstrapped from an abundant language, which is a more practical approach \cite{sitaram2013bootstrapping,muthukumar2014automatic}.

\begin{figure}
  \centering
  \begin{tabular}{cc}
  \includegraphics[width=0.45\linewidth]{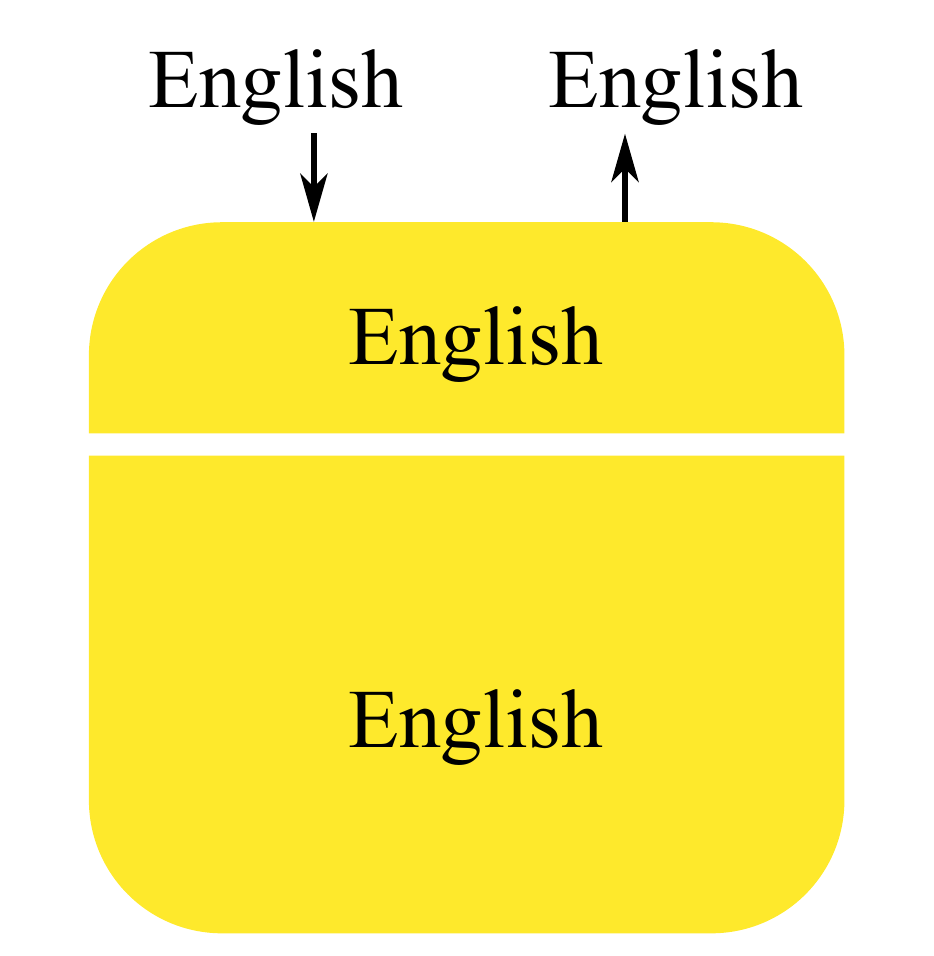} & 
  \includegraphics[width=0.45\linewidth]{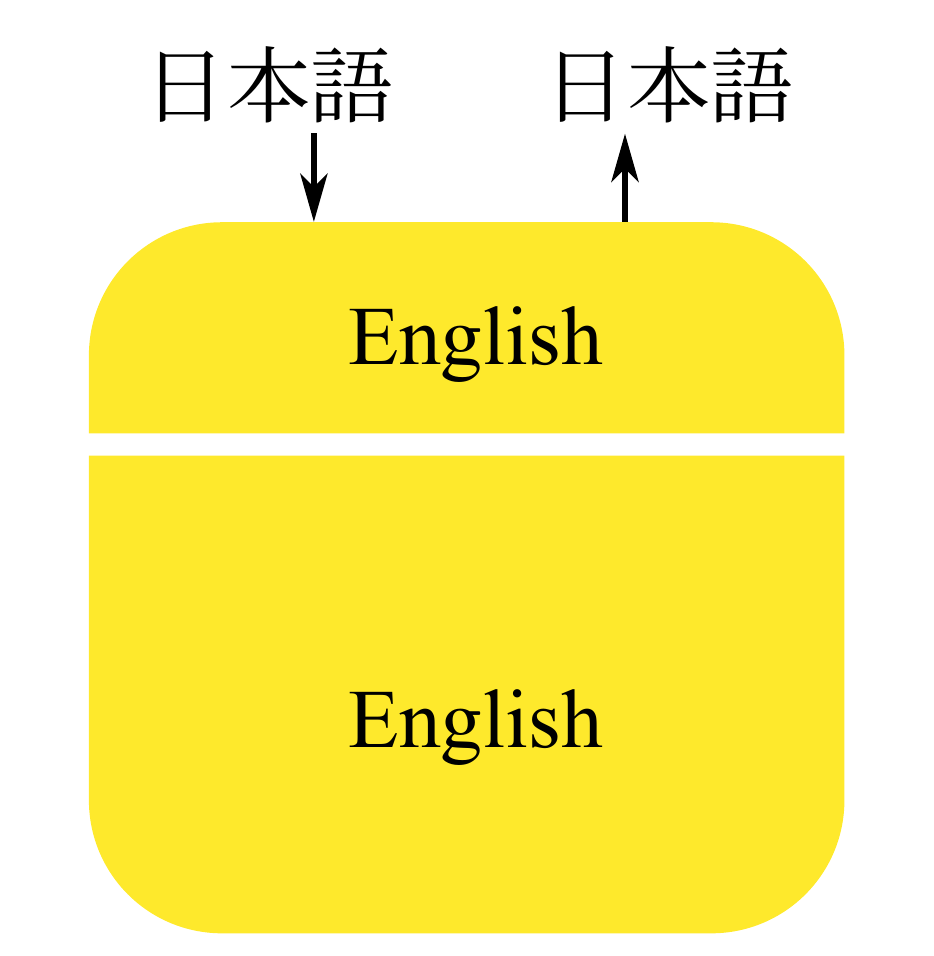} \\
  (a) EE-E & (b) EE-J \\
  \\
  \includegraphics[width=0.45\linewidth]{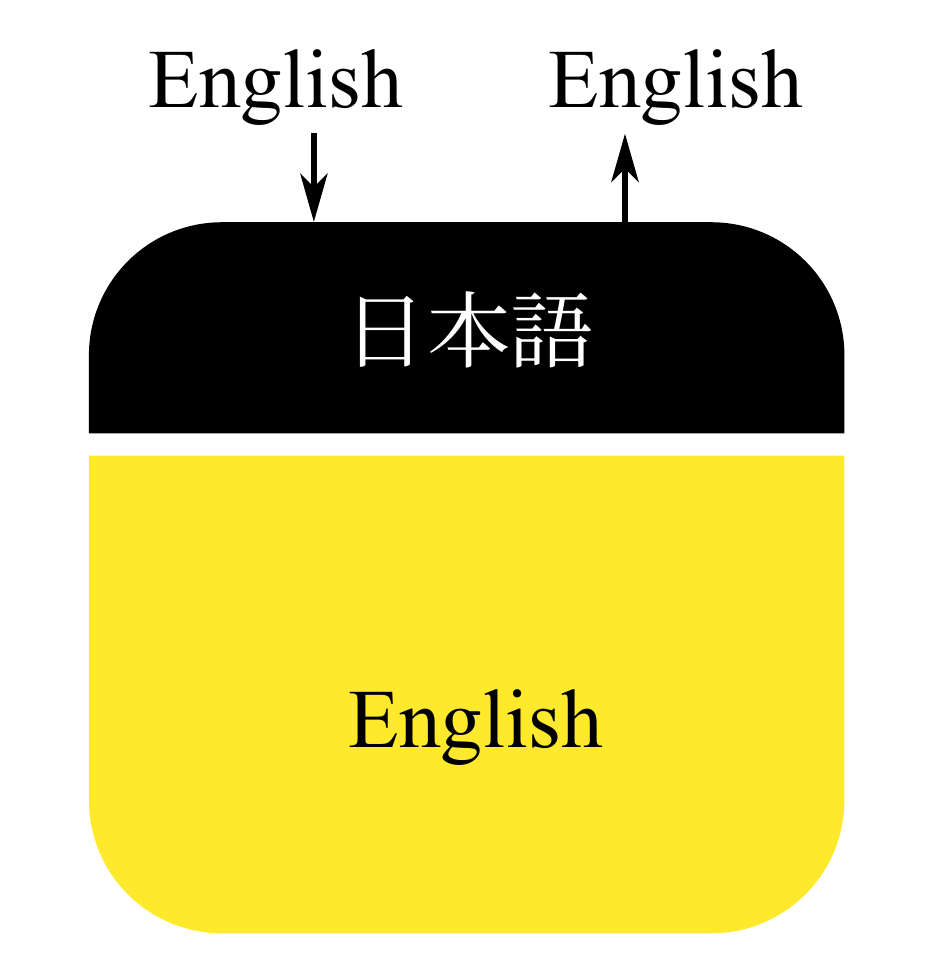} &
  \includegraphics[width=0.45\linewidth]{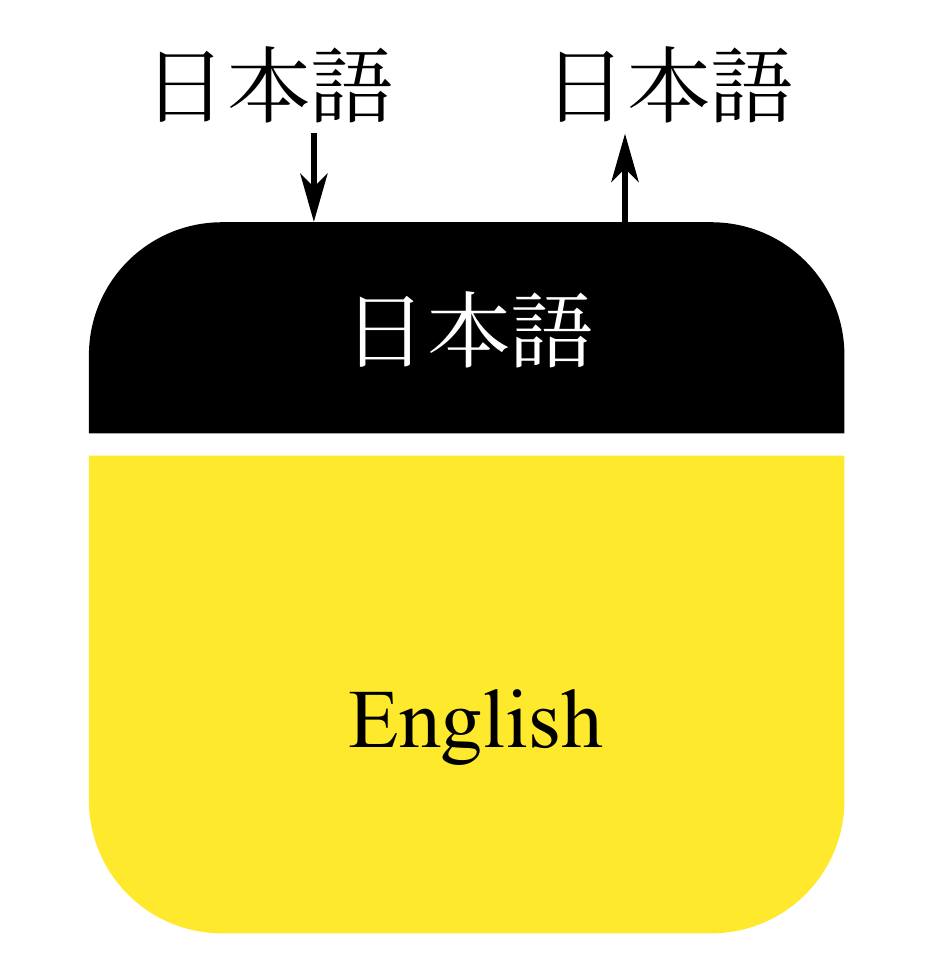} \\
  (c) EJ-E & (d) EJ-J \\
  \end{tabular}
  \caption{Different scenarios of using a non-parallel voice conversion system with unseen language. In this paper, English plays the role of the abundant language (seen) while Japanese plays the role of the low-resource language (unseen).}
\label{fig:lang}
\vspace{-4mm}
\end{figure}

\section{Experiment}
\label{sec:experiment}

\subsection{Dataset}
We used 25.6 hours of transcribed speech from 72 English speakers to train the initial speaker-adaptive TTS model and the initial SI WaveNet vocoder. The data is a subset of the VCTK speech corpus \cite{veaux2017superseded}.
To validate the proposed method, we reenact the SPOKE task of the Voice Conversion Challenge 2018 (VCC2018) \cite{lorenzo2018voice}.
We build the VC system for 4 target speakers (2 males and 2 females) using 81 utterances per person. We then evaluate the SPOKE task using the speech of 4 source speakers (2 males and 2 females); each speaker contribute 35 utterances. Even though the task provided training data for the source speakers, we did not use the training data in our experiments as our model can convert speech of an arbitrary source speaker.

To test the performance of the proposed system in the context of unseen language as described in Section \ref{sec:unseen}, we use 2 in-house bilingual speakers (1 male and 1 female) as the new targets speakers. These two target speakers can speaker English and Japanese at almost native level. 400 utterances per speaker per language are used as training data; 10 more utterances are used for validation. For evaluation, we reuse 2 source speakers (1 male and 1 female) from VCC2018 as the English source speakers and add 2 native Japanese speakers (1 male and 1 female) as the Japanese source speakers. Each source speaker contributes 35 utterances for evaluation.
While Japanese plays the role of the low-resource in our experiments, to provide a good reference, we train an additional system in which Japanese is the abundant language and then use it as the upper bound. This intra-lingual VC system for Japanese, namely JJ-J, is pretrained on 44.9 hours of transcribed speech of 235 native speakers and then adapted to the Japanese training speech of the two bilingual target.

\begin{figure}
  \centering
  \includegraphics[width=0.9\linewidth]{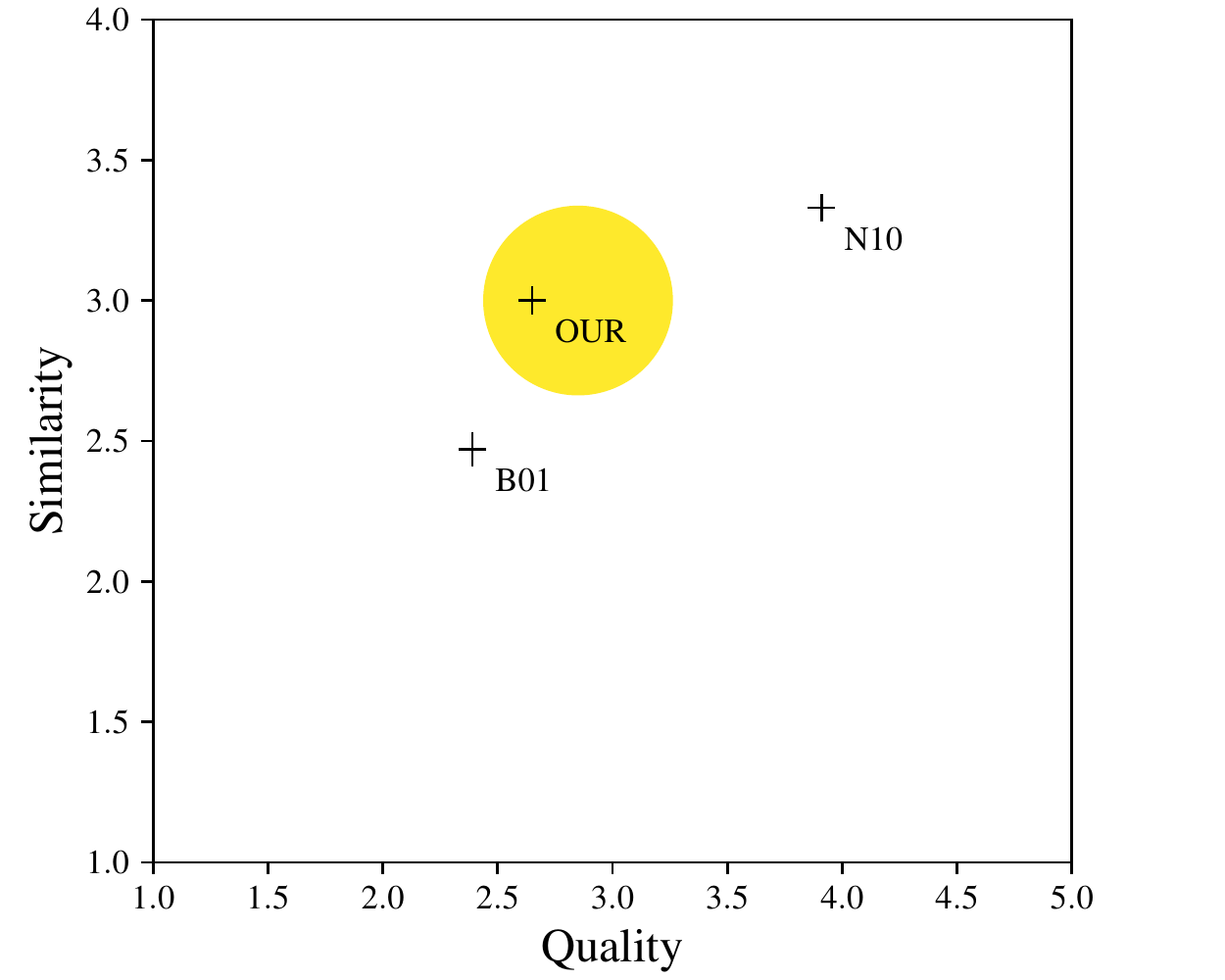}
  \vspace{-3mm}
  \caption{Subjective results for the SPOKE task of the Voice Conversion Challenge 2018. The lines indicate 95\% confidence interval with all results are statistically significant.}
\label{fig:vccspoke}
\vspace{-4mm}
\end{figure}

\subsection{Model configuration}

Our configuration imitates the setup described in \cite{luong2019unified} as closely as possible. The linguistic feature is a 367-dimensional vector contained various information that is deemed to be useful for English speech synthesis. The acoustic feature is the 80-dimensional mel-spectrogram. One difference compared with \cite{luong2019unified} is that all speech data is resampled to 22050 Hz (which is the sampling rate of the VCC2018 dataset) before it is used to extract acoustic features. WaveNet vocoder is also trained on the 22050 Hz waveform that has been quantized using 10-bit u-law.

\begin{table}[t]
    \caption{Detailed subjective results for SPOKE task.}
    \centering
    \vspace{1mm}
    (a) Quality \\
    \vspace{1mm}
    \scalebox{0.9}{
    \begin{tabular}{lrrrrr}
        \hline
        & F-F & F-M & M-F & M-M & ALL\\ \hline
        N10 & 3.91 & 3.96 & 3.85 & 3.93 & 3.91 \\ \hline
        B01 & 3.10 & 2.12 & 1.84 & 2.49 & 2.39 \\ \hline
        OUR & 2.72 & 2.77 & 2.41 & 2.68 & 2.65 \\ \hline
    \end{tabular}}\\

    \vspace{2mm}
    (b) Similarity \\
    \vspace{1mm}
    \centering
    \scalebox{0.9}{
    \begin{tabular}{lrrrrr}
        \hline
        & F-F & F-M & M-F & M-M & All\\ \hline
        N10 & 3.18 & 3.55 & 3.14 & 3.48 & 3.33 \\ \hline
        B01 & 2.74 & 2.87 & 2.21 & 2.04 & 2.47 \\ \hline
        OUR & 2.92 & 3.13 & 2.70 & 3.26 & 3.00 \\ \hline
    \end{tabular}}\\
    \label{tab:detailsim}
    \vspace{-4mm}
\end{table}

The neural network used for the speaker-adaptive acoustic model has the same structure as in \cite{luong2019unified} with the size of the LLE set to 64. Speaker bias is placed at layers B5, B6, B7, and B8 (Fig. 1 in \cite{luong2019unified}) in the form of a one-hot-vector to represent speakers in the training set. The initial multimodal architecture is trained using the loss function defined in Equation \ref{equa:train} with $\beta$ set to 0.25. The SI WaveNet vocoder contained 40 dilated causal layers and it is conditioned on a natural mel-spectrogram. The WaveNet vocoder is fine-tuned for each target speaker using their respective available training speech data.

\begin{figure}[t]
  \centering
  \includegraphics[width=0.9\linewidth]{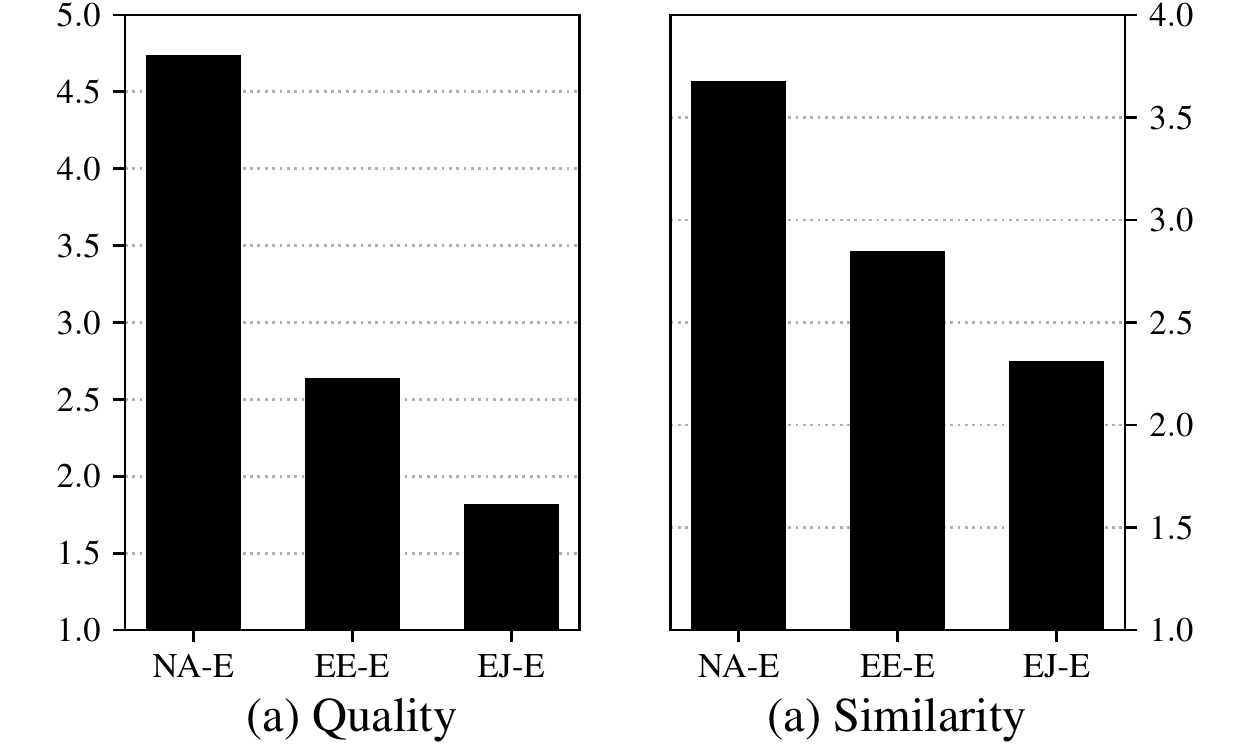}
  \vspace{-4mm}
  \caption{Subjective results for cross-language speaker adaptation task. All results are statistically significant.}
\label{fig:adapt}
\vspace{-4mm}
\end{figure}

\section{Evaluation and Discussion}
\label{sec:evaluation}

\subsection{Voice conversion challenge 2018 SPOKE task}
\label{subsec:eval-vcc}
We follow the guidelines of VCC2018 and build systems for the 4 target speakers of the SPOKE task and converting evaluation utterances of the 4 source speakers.
We compare our system (OUR) with B01 \cite{kobayashi2018sprocket}, which is the baseline of VCC2018, and N10 \cite{liu2018wavenet}, which is the best system based on the subjective test. Instead of reproducing the experiments for B01 and N10, we use the utterances submitted by the participants to ensure the highest quality. We conduct the listening test to judge the quality and similarity of utterances generated by the OUR, B01 and N10 systems. A total of 28 native English speakers participated in our test, most of them did approximately 10 sessions. Each session is prepared to contain generated utterances of every target speaker from every system. In summary, each system is judged 1088 times for quality and another 1088 times for similarity. In the quality question, the participant is asked to judge the quality of the represented speech sample using a 5-point scale mean-opinion score. In the similarity question, the participant is asked to judge the similarity between the utterance generated by one VC system and a natural utterance spoken by the target speaker in a 4-point scale. The setup is the same as in VCC2018 \cite{lorenzo2018voice}.

The general results can be seen in Fig. \ref{fig:vccspoke}. While it is not as good as the best system N10, our system is slightly better than the baseline B01 in terms of quality and significantly better in terms of speaker similarity. One should note that B01 is a strong baseline, it is ranked 3rd in quality and 6th in speaker similarity measurements at the VCC2018 \cite{lorenzo2018voice}. This validated our framework for VC. More detailed results can be found in Table \ref{tab:detailsim}, which shows consistent performance of our system across same-gender and cross-gender pairs.

\subsection{Cross-language speaker adaptation}
\label{subsec:eval-unseen-adapt}

Next, we evaluate our system in the scenario of cross-language speaker adaptation (EJ-E). For this task we develop a system using the speech data of the bilingual target speakers. We compare EJ-E with EE-E, which is the standard intra-language of English, and a reference system NA-E, which is of hold-out natural English utterances.

The native English speakers that participated in the survey for previous task are asked to evaluate the cross-language speaker adaptation task as well. In summary, each scenario is judged 544 times for quality and another 544 times for similarity.
The quality and similarity results are illustrated in Fig. \ref{fig:adapt} and are statistically significant between all scenarios. The cross-language speaker adaptation EJ-E is generally worse than EE-E as one would expect.

\begin{figure}[t]
  \centering
  \includegraphics[width=0.9\linewidth]{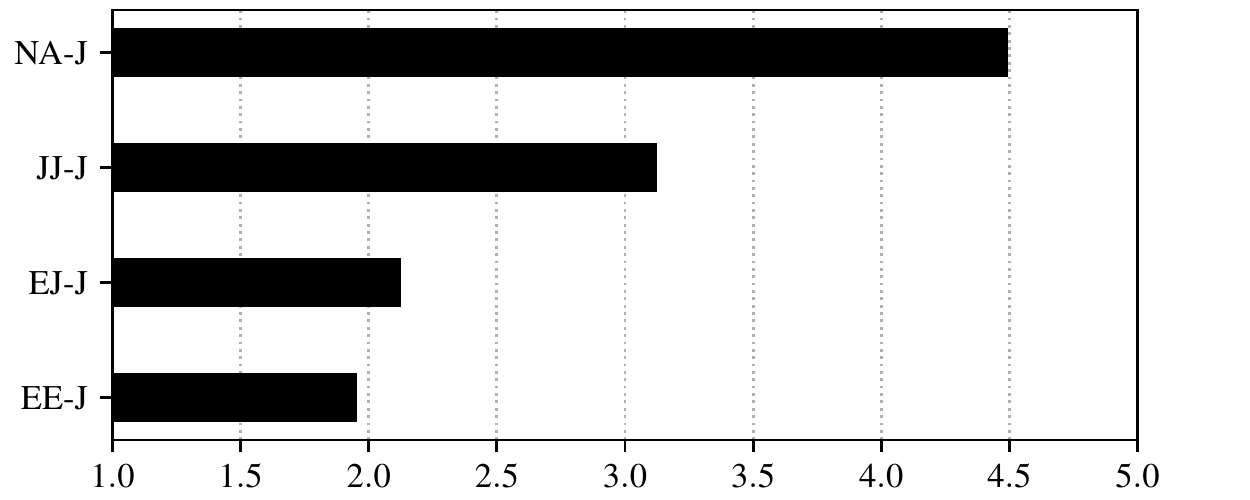}
  \vspace{-4mm}
  \caption{Quality evaluations for low-resource language voice conversion. All results are statistical significant.}
\label{fig:crossqua}
\vspace{-5mm}
\end{figure}

\subsection{Voice conversion for low-resource language}
\label{subsec:eval-unseen-convert}

Finally, we test the performance of our system when using it to convert speech of low-resource language. This consists of the EE-J and EJ-J scenarios. We use JJ-J, the abundant language scenario of Japanese, as the upper-bound and NA-J, the hold-out natural Japanese utterances, as the reference.
We prepare a similar listening test to that of previous tasks. A total 148 native Japanese participated in our test; no participant is allowed to do more than 10 sessions. Each session is prepared to contain all source and target pairs. In summary, each scenario is judged 2800 times for quality and another 2800 times for similarity except the reference natural speech, which is only judged 1400 times for each measurement.

\begin{figure}[t]
  \centering
  \includegraphics[width=0.9\linewidth]{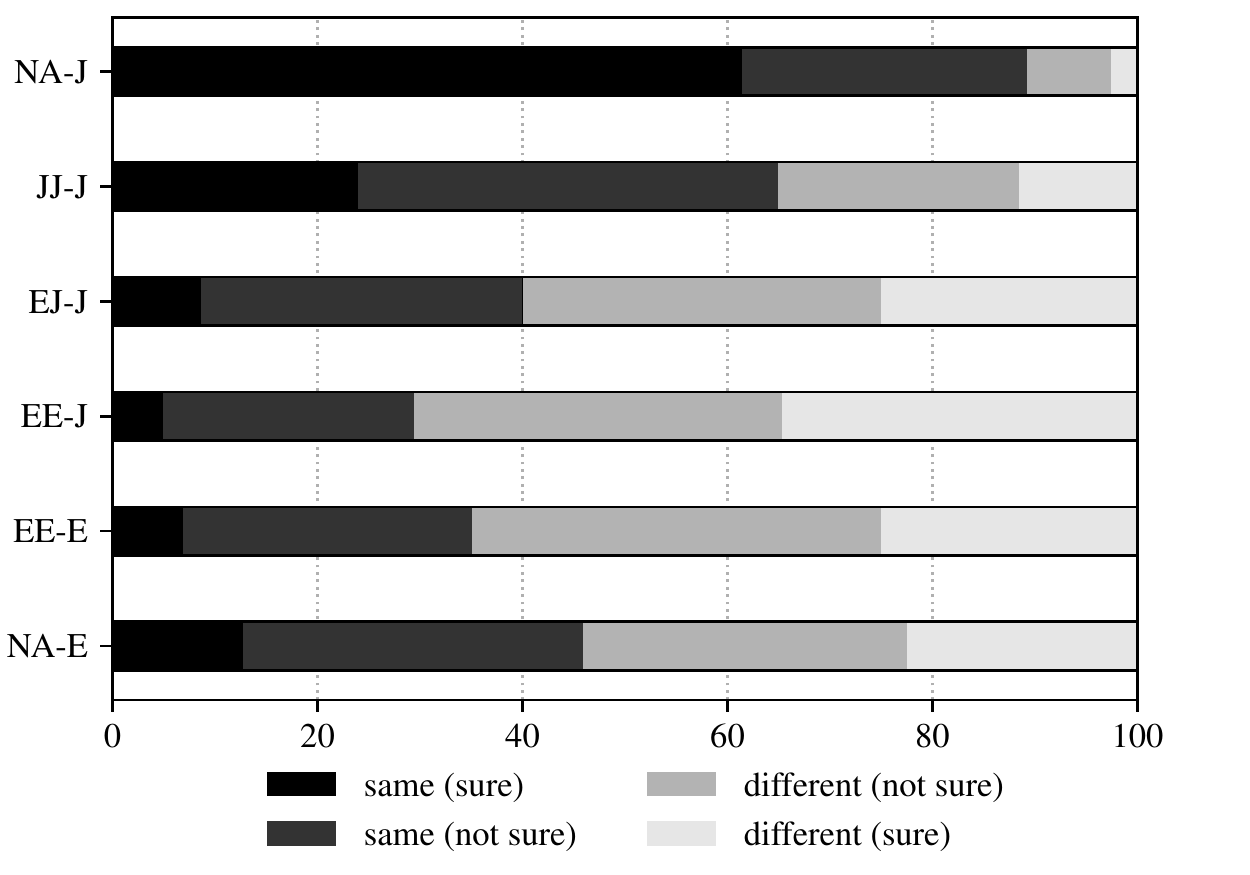}
  \vspace{-5mm}
  \caption{Similarity evaluations for low-resource language voice conversion.}
\label{fig:crosssim}
\vspace{-4mm}
\end{figure}

The result of the quality test is illustrated in Fig. \ref{fig:crossqua}. Our standard Japanese system JJ-J achieved a relatively better score than the English counterpart EE-E, although technically they should not be compared directly. The cross-language converting scenarios, EJ-J and EE-J, are a lot worse than the standard scenario JJ-J; as expected EJ-J has slightly better score than EE-J. The result for the similarity test is illustrated in Fig. \ref{fig:crosssim}. For the four systems that convert Japanese speech, their similarity result trend is the same as their quality result trend. For the similarity test, we also included two extra systems that produce English speech, EE-E and NA-E. In these two cases the listener would listen to an English utterance contributed by EE-E or NA-E and a natural Japanese utterance of the same speaker (as the target speakers are bilingual) and judge their similarity. Interestingly the similarity result of NA-E is a lot worse than that of NA-J even though they both contain natural speech spoken by the same speaker in real life. This suggests that utterances spoken by the same speaker in different languages might not have consistent characteristic. Or problem might be that is it difficult for listeners to evaluate speaker similarity in cross-language scenario, especially when they are not fluent in both.

Even though the performance in the unseen language scenarios is not as good as the standard scenario, the subjective results have shown the ability of using our system for inter-language tasks and establishing a preliminary result as well as a solid baseline for future improvements\footnote{Speech samples are available at \url{https://nii-yamagishilab.github.io/sample-vc-bootstrapping-tts/}}.

\section{Conclusion}
\vspace{-1mm}
\label{sec:conclusion}
In this paper, we have presented a methodology to bootstrap a VC system from a pretrained speaker-adaptive TTS model. By transferring knowledge learned previously by the TTS model, we were able to significantly lower the data requirements for building the VC system. This in turn allows the system to operate in a low-resource unseen language.
The subjective results show that our VC system achieves competitive performance compared with existing methods. Moreover, it can also be used for cross-language voice conversion and cross-language speaker adaptation.
While the performance in these unseen language scenarios are not as good, all experiments in this work are conducted under the assumption of minimal available resources. Our future work includes taking advantage of the available additional resources such as a multi-lingual corpus \cite{zhou2019cross,tu2019end} to further improve the robustness of the VC system.

\bibliographystyle{IEEEbib}
\bibliography{main}

\begin{thebibliography}{10}

\bibitem{toda2007voice}
Tomoki Toda, Alan~W Black, and Keiichi Tokuda,
\newblock ``Voice conversion based on maximum-likelihood estimation of spectral
  parameter trajectory,''
\newblock {\em {IEEE} Trans. Audio, Speech, Language Process.}, vol. 15, no. 8,
  pp. 2222--2235, 2007.

\bibitem{taylor2009text}
Paul Taylor,
\newblock {\em Text-to-speech synthesis},
\newblock Cambridge university press, 2009.

\bibitem{abe1990voice}
Masanobu Abe, Satoshi Nakamura, Kiyohiro Shikano, and Hisao Kuwabara,
\newblock ``Voice conversion through vector quantization,''
\newblock {\em J. Acoust. Soc. Jpn. (E)}, vol. 11, no. 2, pp. 71--76, 1990.

\bibitem{takashima2013exemplar}
Ryoichi Takashima, Tetsuya Takiguchi, and Yasuo Ariki,
\newblock ``Exemplar-based voice conversion using sparse representation in
  noisy environments,''
\newblock {\em IEICE Transactions on Fundamentals of Electronics,
  Communications and Computer Sciences}, vol. 96, no. 10, pp. 1946--1953, 2013.

\bibitem{chen2014voice}
Ling-Hui Chen, Zhen-Hua Ling, Li-Juan Liu, and Li-Rong Dai,
\newblock ``Voice conversion using deep neural networks with layer-wise
  generative training,''
\newblock {\em IEEE/ACM Transactions on Audio, Speech and Language Processing
  (TASLP)}, vol. 22, no. 12, pp. 1859--1872, 2014.

\bibitem{erro2009inca}
Daniel Erro, Asunci{\'o}n Moreno, and Antonio Bonafonte,
\newblock ``{INCA} algorithm for training voice conversion systems from
  nonparallel corpora,''
\newblock {\em {IEEE} Trans. Audio, Speech, Language Process.}, vol. 18, no. 5,
  pp. 944--953, 2009.

\bibitem{chen2003voice}
Yining Chen, Min Chu, Eric Chang, Jia Liu, and Runsheng Liu,
\newblock ``Voice conversion with smoothed gmm and map adaptation,''
\newblock in {\em Proc. EUROSPEECH}, 2003, pp. 2413--2416.

\bibitem{lee2006map}
Chung-Han Lee and Chung-Hsien Wu,
\newblock ``Map-based adaptation for speech conversion using adaptation data
  selection and non-parallel training,''
\newblock in {\em Proc. INTERSPEECH}, 2006, pp. 2254--2257.

\bibitem{toda2006eigenvoice}
Tomoki Toda, Yamato Ohtani, and Kiyohiro Shikano,
\newblock ``Eigenvoice conversion based on gaussian mixture model,''
\newblock in {\em Proc. INTERSPEECH}, 2006, pp. 2446--2449.

\bibitem{sun2016phonetic}
Lifa Sun, Kun Li, Hao Wang, Shiyin Kang, and Helen Meng,
\newblock ``Phonetic posteriorgrams for many-to-one voice conversion without
  parallel data training,''
\newblock in {\em 2016 IEEE International Conference on Multimedia and Expo
  (ICME)}. IEEE, 2016, pp. 1--6.

\bibitem{miyoshi2017voice}
Hiroyuki Miyoshi, Yuki Saito, Shinnosuke Takamichi, and Hiroshi Saruwatari,
\newblock ``Voice conversion using sequence-to-sequence learning of context
  posterior probabilities,''
\newblock {\em Proc. INTERSPEECH}, pp. 1268--1272, 2017.

\bibitem{hsu2016voice}
Chin-Cheng Hsu, Hsin-Te Hwang, Yi-Chiao Wu, Yu~Tsao, and Hsin-Min Wang,
\newblock ``Voice conversion from non-parallel corpora using variational
  auto-encoder,''
\newblock in {\em Proc. APSIPA}. IEEE, 2016, pp. 1--6.

\bibitem{kaneko2017parallel}
Takuhiro Kaneko and Hirokazu Kameoka,
\newblock ``Parallel-data-free voice conversion using cycle-consistent
  adversarial networks,''
\newblock {\em arXiv preprint arXiv:1711.11293}, 2017.

\bibitem{fang2018high}
Fuming Fang, Junichi Yamagishi, Isao Echizen, and Jaime Lorenzo-Trueba,
\newblock ``High-quality nonparallel voice conversion based on cycle-consistent
  adversarial network,''
\newblock in {\em 2018 IEEE International Conference on Acoustics, Speech and
  Signal Processing (ICASSP)}. IEEE, 2018, pp. 5279--5283.

\bibitem{tjandra2019vqvae}
Andros Tjandra, Berrak Sisman, Mingyang Zhang, Sakriani Sakti, Haizhou Li, and
  Satoshi Nakamura,
\newblock ``Vqvae unsupervised unit discovery and multi-scale code2spec
  inverter for zerospeech challenge 2019,''
\newblock {\em arXiv preprint arXiv:1905.11449}, 2019.

\bibitem{yeh2018rhythm}
Cheng-chieh Yeh, Po-chun Hsu, Ju-chieh Chou, Hung-yi Lee, and Lin-shan Lee,
\newblock ``Rhythm-flexible voice conversion without parallel data using
  cycle-gan over phoneme posteriorgram sequences,''
\newblock in {\em Proc. SLT}, 2018, pp. 274--281.

\bibitem{tanaka2019atts2s}
Kou Tanaka, Hirokazu Kameoka, Takuhiro Kaneko, and Nobukatsu Hojo,
\newblock ``Atts2s-vc: Sequence-to-sequence voice conversion with attention and
  context preservation mechanisms,''
\newblock in {\em Proc. ICASSP}, 2019, pp. 6805--6809.

\bibitem{zhang2019non}
Jing-Xuan Zhang, Zhen-Hua Ling, and Li-Rong Dai,
\newblock ``Non-parallel sequence-to-sequence voice conversion with
  disentangled linguistic and speaker representations,''
\newblock {\em arXiv preprint arXiv:1906.10508}, 2019.

\bibitem{luong2019unified}
Hieu-Thi Luong and Junichi Yamagishi,
\newblock ``A unified speaker adaptation method for speech synthesis using
  transcribed and untranscribed speech with backpropagation,''
\newblock {\em arXiv preprint arXiv:1906.07414}, 2019.

\bibitem{jia2018transfer}
Ye~Jia, Yu~Zhang, Ron~J Weiss, Quan Wang, Jonathan Shen, Fei Ren, Zhifeng Chen,
  Patrick Nguyen, Ruoming Pang, Ignacio~Lopez Moreno, and Yonghui Wu,
\newblock ``Transfer learning from speaker verification to multispeaker
  text-to-speech synthesis,''
\newblock {\em arXiv preprint arXiv:1806.04558}, 2018.

\bibitem{nachmani2018fitting}
Eliya Nachmani, Adam Polyak, Yaniv Taigman, and Lior Wolf,
\newblock ``Fitting new speakers based on a short untranscribed sample,''
\newblock {\em arXiv preprint arXiv:1802.06984}, 2018.

\bibitem{stylianou1998continuous}
Yannis Stylianou, Olivier Capp{\'e}, and Eric Moulines,
\newblock ``Continuous probabilistic transform for voice conversion,''
\newblock {\em IEEE Transactions on speech and audio processing}, vol. 6, no.
  2, pp. 131--142, 1998.

\bibitem{kameoka2018acvae}
Hirokazu Kameoka, Takuhiro Kaneko, Kou Tanaka, and Nobukatsu Hojo,
\newblock ``Acvae-vc: Non-parallel many-to-many voice conversion with auxiliary
  classifier variational autoencoder,''
\newblock {\em arXiv preprint arXiv:1808.05092}, 2018.

\bibitem{zhang2019joint}
Mingyang Zhang, Xin Wang, Fuming Fang, Haizhou Li, and Junichi Yamagishi,
\newblock ``Joint training framework for text-to-speech and voice conversion
  using multi-source tacotron and wavenet,''
\newblock {\em arXiv preprint arXiv:1903.12389}, 2019.

\bibitem{sun2016personalized}
Lifa Sun, Hao Wang, Shiyin Kang, Kun Li, and Helen~M Meng,
\newblock ``Personalized, cross-lingual tts using phonetic posteriorgrams.,''
\newblock in {\em Proc. INTERSPEECH}, 2016, pp. 322--326.

\bibitem{abe1991statistical}
Masanobu Abe, Kiyohiro Shikano, and Hisao Kuwabara,
\newblock ``Statistical analysis of bilingual speaker’s speech for
  cross-language voice conversion,''
\newblock {\em J. Acoust. Soc. Am.}, vol. 90, no. 1, pp. 76--82, 1991.

\bibitem{mashimo2001evaluation}
Mikiko Mashimo, Tomoki Toda, Kiyohiro Shikano, and Nick Campbell,
\newblock ``Evaluation of cross-language voice conversion based on {GMM} and
  {S}traight,''
\newblock in {\em Proc. EUROSPEECH}, 2001, pp. 361--364.

\bibitem{zhou2019cross}
Yi~Zhou, Xiaohai Tian, Haihua Xu, Rohan~Kumar Das, and Haizhou Li,
\newblock ``Cross-lingual voice conversion with bilingual phonetic
  posteriorgram and average modeling,''
\newblock in {\em Proc. ICASSP}, 2019, pp. 6790--6794.

\bibitem{rallabandi2019approach}
Sai~Sirisha Rallabandi and Suryakanth~V Gangashetty,
\newblock ``An approach to cross-lingual voice conversion,''
\newblock in {\em Proc. IJCNN}, 2019.

\bibitem{wu2009state}
Yi-Jian Wu, Yoshihiko Nankaku, and Keiichi Tokuda,
\newblock ``State mapping based method for cross-lingual speaker adaptation in
  hmm-based speech synthesis,''
\newblock in {\em Proc. INTERSPEECH}, 2009, pp. 528--531.

\bibitem{dunbar2019zero}
Ewan Dunbar, Robin Algayres, Julien Karadayi, Mathieu Bernard, Juan Benjumea,
  Xuan-Nga Cao, Lucie Miskic, Charlotte Dugrain, Lucas Ondel, Alan~W. Black,
  Laurent Besacier, Sakriani Sakti, and Emmanuel Dupoux,
\newblock ``The zero resource speech challenge 2019: {TTS} without {T},''
\newblock {\em arXiv preprint arXiv:1904.11469}, 2019.

\bibitem{feng2019combining}
Siyuan Feng, Tan Lee, and Zhiyuan Peng,
\newblock ``Combining adversarial training and disentangled speech
  representation for robust zero-resource subword modeling,''
\newblock {\em arXiv preprint arXiv:1906.07234}, 2019.

\bibitem{liu2019unsupervised}
Andy~T Liu, Po-chun Hsu, and Hung-yi Lee,
\newblock ``Unsupervised end-to-end learning of discrete linguistic units for
  voice conversion,''
\newblock {\em arXiv preprint arXiv:1905.11563}, 2019.

\bibitem{eloff2019unsupervised}
Ryan Eloff, Andr{\'e} Nortje, Benjamin van Niekerk, Avashna Govender, Leanne
  Nortje, Arnu Pretorius, Elan Van~Biljon, Ewald van~der Westhuizen, Lisa van
  Staden, and Herman Kamper,
\newblock ``Unsupervised acoustic unit discovery for speech synthesis using
  discrete latent-variable neural networks,''
\newblock {\em arXiv preprint arXiv:1904.07556}, 2019.

\bibitem{sitaram2013bootstrapping}
Sunayana Sitaram, Sukhada Palkar, Yun-Nung Chen, Alok Parlikar, and Alan~W
  Black,
\newblock ``Bootstrapping text-to-speech for speech processing in languages
  without an orthography,''
\newblock in {\em Proc. ICASSP}, 2013, pp. 7992--7996.

\bibitem{muthukumar2014automatic}
Prasanna~Kumar Muthukumar and Alan~W Black,
\newblock ``Automatic discovery of a phonetic inventory for unwritten languages
  for statistical speech synthesis,''
\newblock in {\em Proc. ICASSP}, 2014, pp. 2594--2598.

\bibitem{veaux2017superseded}
Christophe Veaux, Junichi Yamagishi, and Kirsten MacDonald,
\newblock ``{CSTR VCTK} corpus: English multi-speaker corpus for {CSTR} voice
  cloning toolkit,''
\newblock 2017,
\newblock http://dx.doi.org/10.7488/ds/1994.

\bibitem{lorenzo2018voice}
Jaime Lorenzo-Trueba, Junichi Yamagishi, Tomoki Toda, Daisuke Saito, Fernando
  Villavicencio, Tomi Kinnunen, and Zhenhua Ling,
\newblock ``The voice conversion challenge 2018: Promoting development of
  parallel and nonparallel methods,''
\newblock in {\em Proc. Odyssey}, 2018, pp. 195--202.

\bibitem{kobayashi2018sprocket}
Kazuhiro Kobayashi and Tomoki Toda,
\newblock ``sprocket: Open-source voice conversion software,''
\newblock in {\em Proc. Odyssey}, 2018, pp. 203--210.

\bibitem{liu2018wavenet}
Li-Juan Liu, Zhen-Hua Ling, Yuan Jiang, Ming Zhou, and Li-Rong Dai,
\newblock ``Wavenet vocoder with limited training data for voice conversion,''
\newblock in {\em Proc. INTERSPEECH}, 2018, pp. 1983--1987.

\bibitem{tu2019end}
Tao Tu, Yuan-Jui Chen, Cheng-chieh Yeh, and Hung-yi Lee,
\newblock ``End-to-end text-to-speech for low-resource languages by
  cross-lingual transfer learning,''
\newblock {\em arXiv preprint arXiv:1904.06508}, 2019.

\end{thebibliography}

\end{document}